\documentclass[11pt,a4paper]{article}
\usepackage[a4paper,margin=18mm,includefoot]{geometry}
\usepackage[T1]{fontenc}
\usepackage[utf8]{inputenc}
\usepackage{CJKutf8}
\DeclareUnicodeCharacter{2212}{\ensuremath{-}}
\usepackage{lmodern}
\usepackage{microtype}
\usepackage{graphicx}
\usepackage{booktabs}
\usepackage{longtable}
\usepackage{array}
\usepackage{calc}
\usepackage{float}
\usepackage{caption}
\usepackage{url}
\usepackage[hidelinks,unicode]{hyperref}
\usepackage{fancyhdr}
\providecommand{\pandocbounded}[1]{#1}
\title{Post-Edit Re-Verification in Simulator-Backed Engineering Agents:\\
A Controlled Comparison of Verification-Cadence Guidance}
\author{Qingchuan Zhu\textsuperscript{1*}, Shuyue Tong\textsuperscript{1}, Pengju Ren\textsuperscript{2}\\
\small \textsuperscript{1} Sinopec Petroleum Engineering Zhongyuan Co., Ltd., No. 33 Zhengguang North Street,\\
\small Zhengdong New District, Zhengzhou, Henan, China\\
\small \textsuperscript{2} Xi'an Jiaotong University, No. 28 Xianning West Road, Xi'an, Shaanxi 710049, P.R. China\\
\small \textsuperscript{*}Corresponding author: Qingchuan Zhu\\
\small Email: \href{mailto:zhuqingchuan@foxmail.com}{zhuqingchuan@foxmail.com}\\
\small ORCID: \href{https://orcid.org/0000-0003-1816-070X}{0000-0003-1816-070X}}
\date{}
\begin{document}
\maketitle
\begin{abstract}Engineering agents that interact with external simulators may need to
coordinate design modification with reacquisition of engineering
evidence for the modified state. We ask whether first post-edit
re-verification changes when explicit verification-cadence guidance is
retained versus omitted while verification-relevant state/facts are held
constant. Cadence-Guided (CG) retained an instruction to request a new
simulation after a substantive modification, whereas Cadence-Omitted
(CO) removed that instruction; neither condition used a hard gate. The
study therefore measures instruction-conditioned post-edit
verification-policy adherence rather than spontaneous recognition that
prior evidence has become stale. Using DWSIM as the simulator backend
and continuous valve-pressure adjustment, five Alibaba/Qwen models were
evaluated on eight synthetic cases; each model--case--condition
combination was executed three times via live API calls, yielding 120
evaluation slots per condition. Re-verification was observed in 94/120
CG slots versus 32/120 CO slots; cadence violations occurred in 26/120
versus 87/120; and bounded final success was reached in 95/120 versus
35/120. qwen3.5-35b-a3b showed minimal re-verification (1/24 in CG and
0/24 in CO) and no final success in either condition. Within this
bounded protocol, explicit post-edit verification-cadence guidance was
associated with more re-verification, fewer cadence violations, and more
frequent bounded final success, supporting the treatment of verification
cadence as an explicit interaction-protocol component.

\end{abstract}

\section{Introduction}\label{introduction}

Large language model (LLM) agents can interleave language-model
reasoning with calls to external tools, APIs, and environments, allowing
task-relevant observations to be generated outside the model context
(Schick et al.~2023; Yao et al.~2023). This pattern is increasingly
relevant to scientific and engineering workflows, where the information
required for the next decision may be produced by a simulator, solver,
calculation engine, or domain-specific validator rather than being
available directly in text (Bran et al.~2024). In such settings, an
agent must coordinate not only what design action to take, but also when
to obtain new external evidence about the resulting design state.

Recent studies have already connected LLM-based agents with
chemical-process and engineering-simulation workflows. Existing systems
can generate or modify process configurations, execute process
simulators, inspect outputs, and iterate over subsequent actions using
platforms such as industrial flowsheet simulators, AVEVA Process
Simulation, IDAES, and Aspen Plus (Tian et al.~2026; Schäfer et
al.~2026; Liang et al.~2026; Zeng et al.~2025; Tan et al.~2026). Some
workflows explicitly prescribe change--simulate--inspect sequences,
showing that repeated simulator interaction can be incorporated into an
agent protocol (Schäfer et al.~2026). Simulator access and iterative
simulation are therefore treated here as established elements of the
surrounding research context. The narrower question examined in this
study is whether post-edit verification behavior changes when explicit
guidance about the timing of re-verification is retained versus omitted
while the same verification-relevant state/facts remain available.

This question arises from an evidence-freshness problem. Let a design be
in state A, with engineering verification producing evidence \(E(A)\).
If the agent then makes a substantive modification and moves the design
to state B, \(E(A)\) remains evidence about the preceding state; it is
not automatically fresh evidence about B. The agent therefore faces two
distinct protocol decisions: how to modify the design and when the
modified design should be verified again. Iterative tool feedback and
model self-correction provide broader context for this interaction
pattern (Yao et al.~2023; Madaan et al.~2023; Shinn et al.~2023), while
engineering verification practice likewise interprets evidence relative
to the system state, inputs, conditions, and requirements being
evaluated (Hirshorn et al.~2017). Related agent-reliability work has
also discussed evidence provenance, freshness checking, verification
opportunities, action gating, and environment state (Sheng et al.~2026).
In this study, evidence freshness is operationalized narrowly: after a
substantive design-state change, a new DWSIM request/observation pair
constitutes fresh evidence only when it is obtained for the modified
state under the prespecified protocol.

We test this distinction by comparing two conditions that expose the
same verification-relevant engineering state/facts while differing in
whether explicit post-edit verification-cadence guidance is retained or
omitted. The primary outcome is first post-edit re-verification: whether
fresh simulator evidence is obtained after the first substantive
modification and before another substantive modification.

The study uses DWSIM as the simulator backend in a bounded
outlet-pressure repair setting and evaluates five Alibaba/Qwen models
across eight synthetic cases with repeated executions. Its contribution
is a controlled measurement of verification cadence as a distinct
interaction-protocol component while verification-relevant state/facts
are held common. The claim concerns observable post-edit verification
behavior in this simulator-backed protocol.

\section{Methods}\label{methods}

\subsection{Scientific question and study
scope}\label{scientific-question-and-study-scope}

The study question was: when the same verification-relevant engineering
state/facts are visible, how does first post-edit re-verification
behavior differ when explicit verification-cadence guidance is retained
versus omitted?

Engineering repair provided the task context, while the evaluated
construct was post-edit verification-policy behavior.

\subsection{Verification-cadence
conditions}\label{verification-cadence-conditions}

The two conditions received the same explicit verification-relevant
engineering state/facts and differed in whether explicit post-edit
verification-cadence guidance was retained or omitted. CG
(Cadence-Guided) retained guidance that, when the preceding
simulator-backed engineering verification had failed, a substantive
design change had subsequently occurred, and
\texttt{verification\_ready=true}, the next action had to be
\texttt{request\_simulation}. CO (Cadence-Omitted) omitted that
guidance; neither condition used a deterministic hard gate.

{
\begin{longtable}[]{@{}
  >{\raggedright\arraybackslash}p{(\linewidth - 6\tabcolsep) * \real{0.2500}}
  >{\raggedright\arraybackslash}p{(\linewidth - 6\tabcolsep) * \real{0.2500}}
  >{\raggedright\arraybackslash}p{(\linewidth - 6\tabcolsep) * \real{0.2500}}
  >{\raggedright\arraybackslash}p{(\linewidth - 6\tabcolsep) * \real{0.2500}}@{}}
\toprule\noalign{}
\begin{minipage}[b]{\linewidth}\raggedright
Condition
\end{minipage} & \begin{minipage}[b]{\linewidth}\raggedright
Explicit state/facts
\end{minipage} & \begin{minipage}[b]{\linewidth}\raggedright
Cadence policy
\end{minipage} & \begin{minipage}[b]{\linewidth}\raggedright
Hard gate
\end{minipage} \\
\midrule\noalign{}
\endhead
\bottomrule\noalign{}
\endlastfoot
CG (Cadence-Guided) & The same explicit state/facts, including
verification readiness and evidence-freshness information & When the
preceding simulator-backed engineering verification had failed, a
substantive design change had occurred, and
\texttt{verification\_ready=true}, the next action was a new simulation
request & None \\
CO (Cadence-Omitted) & The same explicit state/facts & The cadence
instruction was removed & None \\
\end{longtable}
}

The displayed state/facts included verification readiness, whether the
design had changed since the last verification, the last verification
result, engineering feedback, and the current verification-protocol
state.

The exact CG-only text in the executed Chinese system prompt comprised
the following two sentences:

\begin{quote}
\begin{CJK*}{UTF8}{gbsn}
如果上一轮真实工程验证失败，此后已经发生实质设计改变，且当前verification\_ready=true，下一次动作必须是request\_simulation以获取这次修改后的新工程事实，不得连续进行更多未经验证的设计修改。
这条规则只规定验证节奏，不提供设备、参数或连接的修正答案；修正内容仍由你根据工程反馈自主决定。
\end{CJK*}
\end{quote}

In English, this states that if the preceding engineering verification
failed, a substantive design change has since occurred, and
\texttt{verification\_ready=true}, the next action must be
\texttt{request\_simulation} to obtain new engineering evidence for the
modified design, without further unverified design changes; it also
states that this rule specifies cadence only and supplies no repair
answer. CO omitted these two sentences. The rest of the prompt exposed
the same verification-relevant state/facts in both conditions.

\begin{figure}
\centering
\pandocbounded{\includegraphics[keepaspectratio]{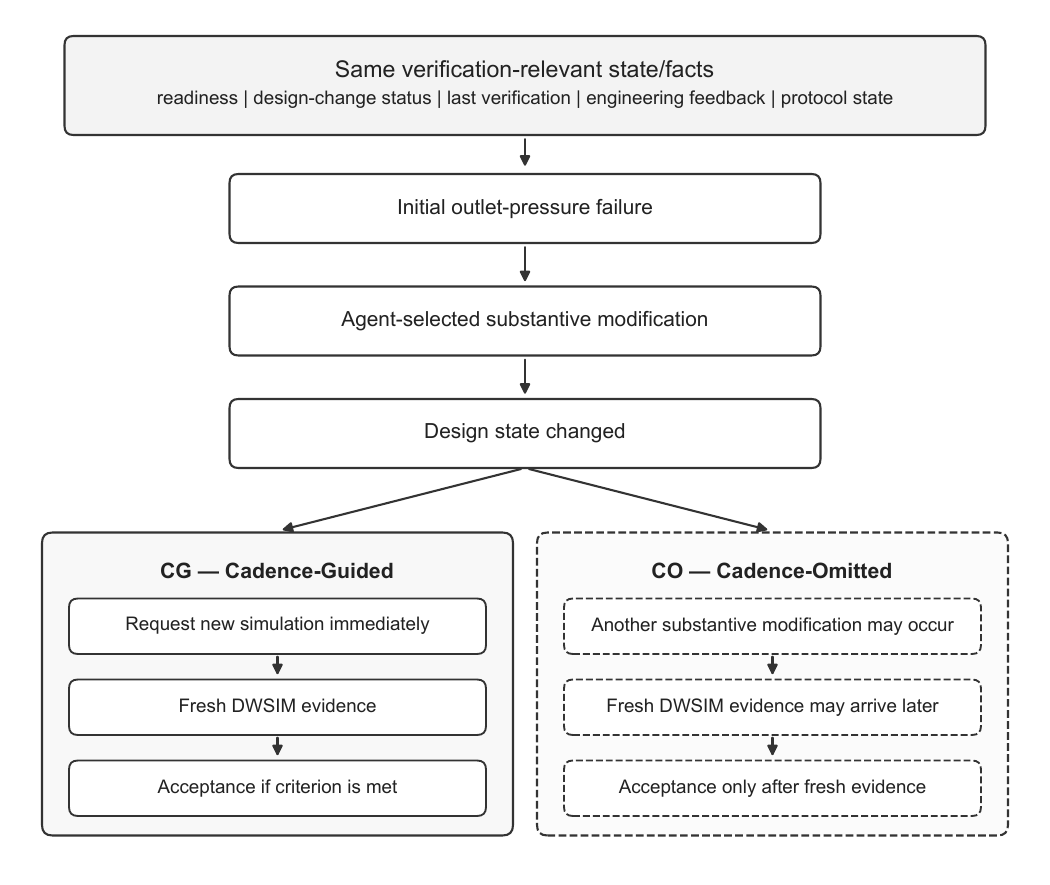}}
\caption{CG/CO post-edit verification protocol. Both
conditions receive the same verification-relevant state/facts and
neither uses a deterministic hard gate. CG retains explicit post-edit
verification-cadence guidance, whereas CO removes that guidance. The
schematic depicts an interaction-protocol contrast, not spontaneous
stale-evidence inference.}
\end{figure}

The cadence instruction specified when fresh verification should be
reacquired; it supplied no repair-specific value, action, parameter
choice, topology change, or deterministic repair rule. Both conditions
received the same engineering requirements, including the
outlet-pressure target. The known feasible repair identified before
evaluation was excluded from the model prompt and context, reward, and
execution advice.

\subsection{Engineering failure
family}\label{engineering-failure-family}

The evaluation used a single outlet-pressure failure family within a
continuous pressure-parameter repair setting. Each case began from a
state that failed the prespecified outlet-pressure acceptance criterion,
and the model selected the next action under a predefined action schema.
The resulting evidence concerns verification-cadence behavior in this
bounded pressure-repair setting.

\subsection{Evaluation design and model
panel}\label{evaluation-design-and-model-panel}

The evaluation used eight synthetic validation cases constructed as a
fixed set of engineering-input combinations rather than sampled from
historical projects or from a population of operating facilities. All
cases used withdrawal mode and the same linear
Feed--separator--valve--heater--filter/separator--Product flowsheet.
They varied standard flow, inlet pressure, target pressure, inlet
temperature, and the initial valve outlet-pressure setting as follows:

{
\begin{longtable}[]{@{}
  >{\raggedright\arraybackslash}p{(\linewidth - 10\tabcolsep) * \real{0.1304}}
  >{\raggedleft\arraybackslash}p{(\linewidth - 10\tabcolsep) * \real{0.1739}}
  >{\raggedleft\arraybackslash}p{(\linewidth - 10\tabcolsep) * \real{0.1739}}
  >{\raggedleft\arraybackslash}p{(\linewidth - 10\tabcolsep) * \real{0.1739}}
  >{\raggedleft\arraybackslash}p{(\linewidth - 10\tabcolsep) * \real{0.1739}}
  >{\raggedleft\arraybackslash}p{(\linewidth - 10\tabcolsep) * \real{0.1739}}@{}}
\toprule\noalign{}
\begin{minipage}[b]{\linewidth}\raggedright
Case
\end{minipage} & \begin{minipage}[b]{\linewidth}\raggedleft
Standard flow (\(10^4\) Nm\(^{3}\)/d)
\end{minipage} & \begin{minipage}[b]{\linewidth}\raggedleft
Inlet pressure (MPa(g))
\end{minipage} & \begin{minipage}[b]{\linewidth}\raggedleft
Target pressure (MPa(g))
\end{minipage} & \begin{minipage}[b]{\linewidth}\raggedleft
Inlet temperature (°C)
\end{minipage} & \begin{minipage}[b]{\linewidth}\raggedleft
Initial valve outlet pressure (MPa(g))
\end{minipage} \\
\midrule\noalign{}
\endhead
\bottomrule\noalign{}
\endlastfoot
\texttt{1} & \texttt{80} & \texttt{10} & \texttt{4.0} & \texttt{18} & \texttt{3.0} \\
\texttt{2} & \texttt{100} & \texttt{11} & \texttt{4.5} & \texttt{20} & \texttt{3.2} \\
\texttt{3} & \texttt{120} & \texttt{12} & \texttt{5.0} & \texttt{22} & \texttt{4.1} \\
\texttt{4} & \texttt{140} & \texttt{13} & \texttt{5.5} & \texttt{24} & \texttt{4.0} \\
\texttt{5} & \texttt{100} & \texttt{14} & \texttt{6.0} & \texttt{26} & \texttt{5.1} \\
\texttt{6} & \texttt{160} & \texttt{12} & \texttt{4.5} & \texttt{28} & \texttt{3.4} \\
\texttt{7} & \texttt{180} & \texttt{15} & \texttt{7.0} & \texttt{30} & \texttt{6.2} \\
\texttt{8} & \texttt{120} & \texttt{16} & \texttt{7.5} & \texttt{32} & \texttt{6.1} \\
\end{longtable}
}

The H-101 heater outlet-temperature setting (35 °C), maximum stage ratio
(3.2), maximum discharge temperature (150 °C), maximum single-stage
throttle drop (100 bar), Joule--Thomson coefficient (0.45 K/bar),
hydrate-temperature margin (3 °C), gas molecular weight (18.5 kg/kmol),
heat-capacity ratio (1.28), compressibility factor (0.9), free-water
evidence flag (false), and available utilities (electric and gas engine)
were held constant. The common input schema also contained
\texttt{cooler\_outlet\_temperature\_deg\_c=35}; this field applied only
if cooling was used, and the corresponding conditional requirement was
inactive because no cooler was present in the evaluated flowsheets. Each
initial valve setting was below its case-specific target, and the
initial state of every case was verified with DWSIM to produce
\texttt{OUTLET\_PRESSURE\_NOT\_MET} before model evaluation.

An \emph{evaluation slot} denotes one model × case × repeat × condition
combination. Each model--case--condition combination was executed three
times as separate live provider calls with identical configured decoding
settings: temperature 0, thinking disabled, JSON-object response format,
and client-side retries disabled. All live API executions were conducted
on 25 August 2026. No per-call model-generation seed was supplied or
exposed by the provider interface; the repeats therefore capture
run-to-run provider/model variation under the same settings rather than
variation in a sampling parameter. Execution order was fixed by a
balanced schedule generated with seed 20260825; that seed controlled
schedule order only, not model generation. The three repeats were nested
within each case, yielding 24 evaluation slots per condition per model
(8 cases × 3 repeats), not 24 independent engineering scenarios.

For Alibaba, five Qwen models were evaluated with 24 CG evaluation slots
and 24 CO evaluation slots per model. This produced 48 evaluation slots
per model, 240 evaluation slots across the panel, and 120 evaluation
slots in each condition.

The Alibaba panel contained the following Qwen models:

{
\begin{longtable}[]{@{}l@{}}
\toprule\noalign{}
Model \\
\midrule\noalign{}
\endhead
\bottomrule\noalign{}
\endlastfoot
qwen3.5-27b \\
qwen3.5-35b-a3b \\
qwen3.5-122b-a10b \\
qwen3.6-27b \\
qwen3.5-397b-a17b \\
\end{longtable}
}

The five-model panel was not treated as a pure dense parameter-scaling
sequence because qwen3.5-397b-a17b was a large mixture-of-experts (MoE)
endpoint. No scaling analysis was performed.

\subsection{Common execution and engineering verification
path}\label{common-execution-and-engineering-verification-path}

Within each model, both conditions used the same requirements, state
representation, legality checker, action schema, simulator path, and
episode budgets; the prompt wording described above constituted the
intended condition-level contrast.

The model selected actions, the legality checker verified action
validity and ranges, the agent framework tracked design state and
evidence freshness, and DWSIM returned engineering observations. This
division separated action selection from legality checking, state
tracking, and engineering verification.

The simulator records identify DWSIM 9.0.5 and the Peng--Robinson (PR)
property package. The valve outlet-pressure setting was passed to the
DWSIM request, and each request/observation pair was linked to the
corresponding design-state fingerprint. The initial condition failed the
outlet-pressure acceptance criterion; a repaired run had to satisfy both
the pressure target and simulator convergence. Solver convergence and
acceptance were therefore treated as distinct checks. Exact simulator
reproduction from the information reported here is limited because a
complete flowsheet configuration and all additional thermodynamic
settings are not specified.

\subsection{Primary outcomes and
reporting}\label{primary-outcomes-and-reporting}

The three primary metrics were prespecified and applied without
redefinition.

{
\begin{longtable}[]{@{}
  >{\raggedright\arraybackslash}p{(\linewidth - 4\tabcolsep) * \real{0.3333}}
  >{\raggedright\arraybackslash}p{(\linewidth - 4\tabcolsep) * \real{0.3333}}
  >{\raggedright\arraybackslash}p{(\linewidth - 4\tabcolsep) * \real{0.3333}}@{}}
\toprule\noalign{}
\begin{minipage}[b]{\linewidth}\raggedright
Metric
\end{minipage} & \begin{minipage}[b]{\linewidth}\raggedright
Operational definition
\end{minipage} & \begin{minipage}[b]{\linewidth}\raggedright
Denominator
\end{minipage} \\
\midrule\noalign{}
\endhead
\bottomrule\noalign{}
\endlastfoot
Reverify & For an episode with at least one substantive modification,
after the first substantive modification and before the next substantive
modification, a new DWSIM request/observation pair was obtained &
Episodes with at least one substantive modification \\
Cadence violation & After the first substantive modification and before
fresh DWSIM evidence, a second accepted substantive modification
occurred & The same episodes with at least one substantive
modification \\
Final success & A successful run with a post-modification fresh DWSIM
request/observation pair and evidence that the acceptance criterion was
met & The 24 evaluation slots in the model block \\
\end{longtable}
}

A substantive modification was an accepted action accompanied by an
actual change in the design state or state fingerprint. A simulation
request was not a substantive modification. These definitions determine
the numerators and denominators reported below and were specified before
the results were assembled.

All 120 evaluation slots in each Alibaba condition contained at least
one substantive modification and were therefore eligible for the
re-verification and cadence-violation denominators.

Because the CG treatment names the required next action and the primary
reverify outcome records that same post-edit event, the estimand is
first post-edit, instruction-conditioned verification-policy adherence.

We report numerator/denominator values at the model level, with each
model contributing 24 evaluation slots per condition. Because the three
repeated executions are nested within eight cases, the 120 CG and 120 CO
evaluation-slot totals are treated as descriptive panel summaries rather
than independent engineering scenarios. Differences are reported as
counts, percentages, and descriptive percentage-point contrasts; no
population-level inferential analysis or hypothesis testing was
performed.

\section{Results}\label{results}

\subsection{Alibaba model-level
results}\label{alibaba-model-level-results}

Model-level results are the primary reporting boundary for the Alibaba
panel:

\begin{figure}
\centering
\pandocbounded{\includegraphics[keepaspectratio]{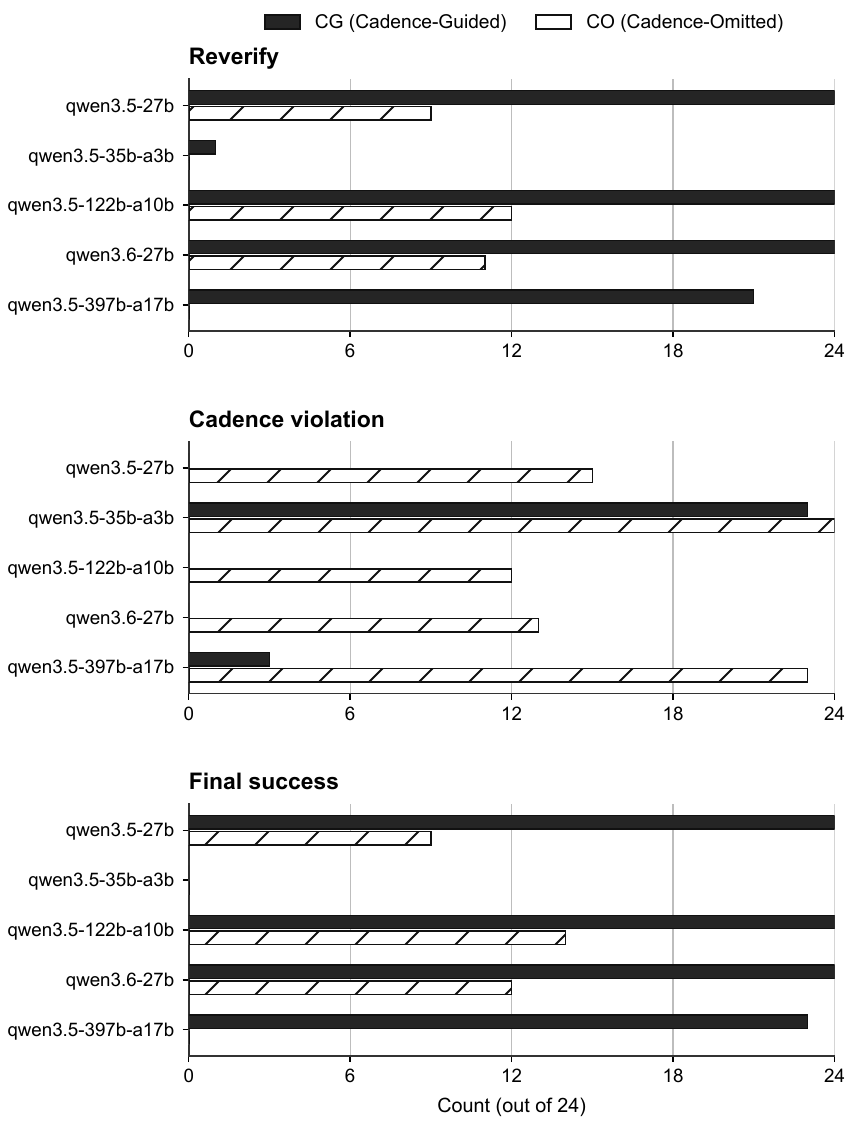}}
\caption{Model-level outcomes. Counts are shown out of 24
evaluation slots per model and condition. The model-level rows, rather
than the 120-slot Alibaba aggregate, are the primary reporting boundary.
qwen3.5-397b-a17b is a large-MoE endpoint rather than a scaling point.}
\end{figure}

{
\begin{longtable}[]{@{}
  >{\raggedright\arraybackslash}p{(\linewidth - 12\tabcolsep) * \real{0.1429}}
  >{\raggedright\arraybackslash}p{(\linewidth - 12\tabcolsep) * \real{0.1429}}
  >{\raggedright\arraybackslash}p{(\linewidth - 12\tabcolsep) * \real{0.1429}}
  >{\raggedright\arraybackslash}p{(\linewidth - 12\tabcolsep) * \real{0.1429}}
  >{\raggedright\arraybackslash}p{(\linewidth - 12\tabcolsep) * \real{0.1429}}
  >{\raggedright\arraybackslash}p{(\linewidth - 12\tabcolsep) * \real{0.1429}}
  >{\raggedright\arraybackslash}p{(\linewidth - 12\tabcolsep) * \real{0.1429}}@{}}
\toprule\noalign{}
\begin{minipage}[b]{\linewidth}\raggedright
Model
\end{minipage} & \begin{minipage}[b]{\linewidth}\raggedright
CG reverify
\end{minipage} & \begin{minipage}[b]{\linewidth}\raggedright
CO reverify
\end{minipage} & \begin{minipage}[b]{\linewidth}\raggedright
CG cadence violation
\end{minipage} & \begin{minipage}[b]{\linewidth}\raggedright
CO cadence violation
\end{minipage} & \begin{minipage}[b]{\linewidth}\raggedright
CG final success
\end{minipage} & \begin{minipage}[b]{\linewidth}\raggedright
CO final success
\end{minipage} \\
\midrule\noalign{}
\endhead
\bottomrule\noalign{}
\endlastfoot
qwen3.5-27b & \texttt{24/24} & \texttt{9/24} & \texttt{0/24} & \texttt{15/24} & \texttt{24/24} & \texttt{9/24} \\
qwen3.5-35b-a3b & \texttt{1/24} & \texttt{0/24} & \texttt{23/24} & \texttt{24/24} & \texttt{0/24} & \texttt{0/24} \\
qwen3.5-122b-a10b & \texttt{24/24} & \texttt{12/24} & \texttt{0/24} & \texttt{12/24} & \texttt{24/24} & \texttt{14/24} \\
qwen3.6-27b & \texttt{24/24} & \texttt{11/24} & \texttt{0/24} & \texttt{13/24} & \texttt{24/24} & \texttt{12/24} \\
qwen3.5-397b-a17b & \texttt{21/24} & \texttt{0/24} & \texttt{3/24} & \texttt{23/24} & \texttt{23/24} & \texttt{0/24} \\
\end{longtable}
}

qwen3.5-27b recorded higher CG values for reverify and final success and
fewer CG cadence violations than CO. qwen3.5-122b-a10b and qwen3.6-27b
showed the same direction across each of the three outcomes.
qwen3.5-397b-a17b also showed the same direction, with 21/24 versus 0/24
for reverify, 3/24 versus 23/24 for cadence violation, and 23/24 versus
0/24 for final success.

qwen3.5-35b-a3b showed minimal re-verification and no final success in
either condition: CG recorded 1/24 reverify, 23/24 cadence violation,
and 0/24 final success, while CO recorded 0/24, 24/24, and 0/24,
respectively.

Across the model rows, CG had a higher reverify count than CO in every
model, including qwen3.5-35b-a3b (1/24 versus 0/24). The other four
models also had higher CG final-success counts and lower CG
cadence-violation counts than CO. Because outcomes varied materially by
model, results are retained at the model level.

\subsection{Descriptive Alibaba panel
aggregate}\label{descriptive-alibaba-panel-aggregate}

Across the Alibaba panel, the descriptive evaluation-slot total for
reverify was 94/120 (78.3\%) in CG versus 32/120 (26.7\%) in CO, a
descriptive CG-minus-CO difference of +51.7 percentage points. Here 120
= 5 models × 8 cases × 3 repeats per condition; it is not 120
independent engineering cases. Cadence violations were recorded for
26/120 CG evaluation slots (21.7\%) and 87/120 CO evaluation slots
(72.5\%), a descriptive difference of −50.8 percentage points. Final
success, defined as the bounded joint within-protocol endpoint requiring
repair, fresh post-modification verification, and acceptance evidence,
was recorded for 95/120 CG evaluation slots (79.2\%) and 35/120 (29.2\%)
in CO, a descriptive difference of +50.0 percentage points.

{
\begin{longtable}[]{@{}
  >{\raggedright\arraybackslash}p{(\linewidth - 6\tabcolsep) * \real{0.2500}}
  >{\raggedright\arraybackslash}p{(\linewidth - 6\tabcolsep) * \real{0.2500}}
  >{\raggedright\arraybackslash}p{(\linewidth - 6\tabcolsep) * \real{0.2500}}
  >{\raggedright\arraybackslash}p{(\linewidth - 6\tabcolsep) * \real{0.2500}}@{}}
\toprule\noalign{}
\begin{minipage}[b]{\linewidth}\raggedright
Primary outcome (descriptive evaluation slots)
\end{minipage} & \begin{minipage}[b]{\linewidth}\raggedright
CG
\end{minipage} & \begin{minipage}[b]{\linewidth}\raggedright
CO
\end{minipage} & \begin{minipage}[b]{\linewidth}\raggedright
Descriptive CG minus CO
\end{minipage} \\
\midrule\noalign{}
\endhead
\bottomrule\noalign{}
\endlastfoot
Reverify & \texttt{94/120} (78.3\%) & \texttt{32/120} (26.7\%) & +51.7 pp \\
Cadence violation & \texttt{26/120} (21.7\%) & \texttt{87/120} (72.5\%) & −50.8 pp \\
Final success & \texttt{95/120} (79.2\%) & \texttt{35/120} (29.2\%) & +50.0 pp \\
\end{longtable}
}

\section{Discussion}\label{discussion}

\subsection{Principal finding}\label{principal-finding}

Within the evaluated simulator-backed pressure-repair protocol, first
post-edit re-verification behavior differed substantially when explicit
verification-cadence guidance was retained versus omitted while the same
verification-relevant state/facts remained visible. Across the
Alibaba/Qwen panel, CG produced higher re-verification counts than CO in
every model row. The other four models also showed fewer cadence
violations and higher bounded final-success counts under CG, whereas
qwen3.5-35b-a3b showed minimal re-verification and no final success in
either condition. At the descriptive panel level, re-verification was
observed in 94/120 CG evaluation slots versus 32/120 CO evaluation
slots, cadence violations in 26/120 versus 87/120, and bounded final
success in 95/120 versus 35/120.

The central interpretation is behavioral: explicit post-edit cadence
guidance was associated with a different observable interaction sequence
between design modification and reacquisition of simulator evidence.
Because CG directly specifies the post-edit simulation request measured
by the primary outcome, the result is best read as
instruction-conditioned tool-use behavior. This supports treating
verification cadence as an explicit part of the interaction protocol for
simulator-backed engineering agents.

\subsection{State exposure and verification cadence as protocol
components}\label{state-exposure-and-verification-cadence-as-protocol-components}

CG and CO received the same verification-relevant state/facts, including
verification readiness, design-change status, the preceding verification
result, engineering feedback, and verification-protocol state. The
controlled difference was whether explicit post-edit cadence guidance
was retained, isolating within the implemented protocol the presence or
omission of an instruction governing when fresh simulator evidence
should be reacquired.

This distinction is useful at the framework level. State exposure and
verification cadence can be implemented as separately configurable
components of an engineering-agent interaction protocol: one represents
information available to the agent, while the other specifies when a new
observation is due after a state-changing action. Related work on agent
reliability has discussed evidence provenance, freshness, verification
opportunities, action gating, and environment state (Sheng et al.~2026).
The present study provides a simulator-backed operationalization of this
distinction.

\subsection{Repair outcome, model heterogeneity, and reporting
unit}\label{repair-outcome-model-heterogeneity-and-reporting-unit}

The model selected the engineering action, the legality checker assessed
action validity, the agent framework tracked the state and evidence
boundary, and DWSIM supplied the resulting observation. Final success is
a bounded joint outcome requiring repair, fresh post-modification
verification, and satisfaction of the acceptance criterion. Because
repair trajectories differed across conditions, this contrast does not
isolate repair capability by itself.

The model-level results also show heterogeneous responses to cadence
guidance. In particular, qwen3.5-35b-a3b recorded 1/24 versus 0/24
re-verification, 23/24 versus 24/24 cadence violations, and 0/24 versus
0/24 final success. The panel was not designed as a controlled
parameter-scaling experiment, so this heterogeneity is reported
descriptively rather than interpreted as a scaling pattern.

Because the three repeated executions are nested within eight cases per
model, the 120 evaluation slots per condition are descriptive panel
totals. Model-level numerator/denominator results remain the primary
reporting boundary.

\subsection{Relation to prior work and framework
implications}\label{relation-to-prior-work-and-framework-implications}

Prior work has demonstrated workflows in which LLM-based agents generate
or modify process configurations, execute simulators, inspect outputs,
and iterate using chemical-process simulation environments (Tian et
al.~2026; Schäfer et al.~2026; Liang et al.~2026; Zeng et al.~2025; Tan
et al.~2026). Schäfer et al.~provide particularly relevant prior work
through a change--simulate--inspect sequence in an agentic
flowsheet-simulation workflow (Schäfer et al.~2026). The present
contribution is the controlled comparison of post-edit verification
cadence while verification-relevant state/facts are held common.

From a framework perspective, three responsibilities can be
distinguished: exposing the current engineering state and provenance of
prior evidence; specifying when fresh verification should be requested
after a state-changing action; and, where required, enforcing that
boundary deterministically. The present experiment compares the first
two, while direct comparison between soft cadence guidance and
deterministic enforcement remains a topic for future work.

\subsection{Limitations and future
work}\label{limitations-and-future-work}

The study is limited to a single outlet-pressure failure family,
continuous valve-pressure repair, eight synthetic cases with three
repeated executions, and a prompt-level intervention. The repeated
observations are not independent engineering scenarios, and no
population-level inferential analysis was performed. Exact simulator
reproduction is also limited because a complete flowsheet configuration
and all additional thermodynamic settings are not reported.

Future work should extend the evaluation to additional failure families,
broader engineering design changes, additional models and providers,
alternative cadence formulations, and direct comparison between soft
guidance and deterministic enforcement.

\section{Conclusions}\label{conclusions}

Within the evaluated simulator-backed pressure-repair protocol, first
post-edit re-verification behavior differed substantially when explicit
verification-cadence guidance was retained versus omitted while the same
verification-relevant state/facts remained visible. Across the
five-model Alibaba/Qwen panel, re-verification was observed in 94/120 CG
evaluation slots versus 32/120 CO evaluation slots, cadence violations
in 26/120 versus 87/120, and bounded final success in 95/120 versus
35/120. The model-level pattern was heterogeneous: qwen3.5-35b-a3b
showed minimal re-verification and no final success in either condition,
while the other four models showed higher final-success counts and fewer
cadence violations under CG than under CO.

These results support treating verification-relevant state exposure and
post-edit verification cadence as separately configurable components of
simulator-backed engineering-agent protocols. The finding is limited to
the evaluated outlet-pressure repair setting and descriptive
repeated-execution panel, and should be tested across broader
engineering tasks, model families, and verification-control mechanisms.

\section*{Declaration of Generative AI and AI-Assisted
Technologies}\label{declaration-of-generative-ai-and-ai-assisted-technologies}

During the preparation of this work, Qingchuan Zhu used OpenAI ChatGPT and
Codex to assist with manuscript organization, language refinement, and
code and document preparation. Qingchuan Zhu reviewed and edited the
resulting material and takes full responsibility for the content of the
manuscript.

Generative AI models evaluated as part of the study are described
separately in the Methods and are not covered by this
manuscript-preparation disclosure.

\section*{Competing Interests}\label{competing-interests}

The authors declare no competing interests.

\section*{References}\label{references}

Bran, Andres M., Sam Cox, Oliver Schilter, et al.~2024. ``Augmenting
Large Language Models with Chemistry Tools.'' \emph{Nature Machine
Intelligence} 6: 525--35. DOI: 10.1038/s42256-024-00832-8; URL:
\url{https://www.nature.com/articles/s42256-024-00832-8}.

Hirshorn, Steven R., Linda D. Voss, and Linda K. Bromley. 2017.
\emph{NASA Systems Engineering Handbook}. NASA/SP-2016-6105 Rev 2.
National Aeronautics and Space Administration. URL:
\url{https://ntrs.nasa.gov/archive/nasa/casi.ntrs.nasa.gov/20170001761.pdf}.

Liang, Jingkang, Niklas Groll, and Gürkan Sin. 2026. ``Large Language
Model Agent for User-Friendly Chemical Process Simulations.''
\emph{Digital Chemical Engineering} 19: 100312. DOI:
10.1016/j.dche.2026.100312; URL:
\url{https://www.sciencedirect.com/science/article/pii/S2772508126000256}.

Madaan, Aman, Niket Tandon, Prakhar Gupta, et al.~2023. ``Self-Refine:
Iterative Refinement with Self-Feedback.'' \emph{Advances in Neural
Information Processing Systems} 36. DOI: 10.52202/075280-2019; URL:
\url{https://papers.neurips.cc/paper_files/paper/2023/hash/91edff07232fb1b55a505a9e9f6c0ff3-Abstract-Conference.html}.

Schäfer, Pascal, Lukas J. Krinke, Martin Wlotzka, and Norbert Asprion.
2026. ``Context Is All You Need: Toward Autonomous Model-Based Process
Design Using Agentic AI in Flowsheet Simulations.'' \emph{AIChE
Journal}, e70488. DOI: 10.1002/aic.70488; URL:
\url{https://aiche.onlinelibrary.wiley.com/doi/10.1002/aic.70488}.

Schick, Timo, J. Dwivedi-Yu, R. Dessi, et al.~2023. ``Toolformer:
Language Models Can Teach Themselves to Use Tools.'' \emph{Advances in
Neural Information Processing Systems} 36. DOI: 10.52202/075280-2997;
URL:
\url{https://proceedings.neurips.cc/paper/2023/hash/d842425e4bf79ba039352da0f658a906-Abstract-Conference.html}.

Sheng, Strick, Ziyue Wang, and Liyi Zhou. 2026. \emph{When Agents
Overtrust Environmental Evidence: An Extensible Agentic Framework for
Benchmarking Evidence-Grounding Defects in LLM Agents}. DOI:
10.48550/arXiv.2605.08828; URL: \url{https://arxiv.org/abs/2605.08828}.

Shinn, Noah, Federico Cassano, Ashwin Gopinath, Karthik Narasimhan, and
Shunyu Yao. 2023. ``Reflexion: Language Agents with Verbal Reinforcement
Learning.'' \emph{Advances in Neural Information Processing Systems} 36.
URL:
\url{https://papers.neurips.cc/paper_files/paper/2023/hash/1b44b878bb782e6954cd888628510e90-Abstract-Conference.html}.

Tan, Sihan, Xiaochi Zhou, Hai Zhou, et al.~2026.
``Reasoning-Agent-Driven Process Simulation, Optimization, Carbon
Accounting and Decarbonization of Distillation.'' \emph{Communications
Engineering} 5: 26. DOI: 10.1038/s44172-025-00583-3; URL:
\url{https://www.nature.com/articles/s44172-025-00583-3}.

Tian, Xufei, Wenli Du, Shaoyi Yang, et al.~2026. ``From Text to
Simulation: A Multi-Agent LLM Workflow for Automated Chemical Process
Design.'' \emph{Proceedings of the AAAI Conference on Artificial
Intelligence} 40: 29705--13. DOI: 10.1609/aaai.v40i35.40215; URL:
\url{https://ojs.aaai.org/index.php/AAAI/article/view/40215}.

Yao, Shunyu, Jeffrey Zhao, Dian Yu, et al.~2023. ``ReAct: Synergizing
Reasoning and Acting in Language Models.'' \emph{International
Conference on Learning Representations}. URL:
\url{https://openreview.net/forum?id=WE_vluYUL-X}.

Zeng, Tong, Srivathsan Badrinarayanan, Janghoon Ock, Cheng-Kai Lai, and
Amir Barati Farimani. 2025. ``LLM-Guided Chemical Process Optimization
with a Multi-Agent Approach.'' \emph{Machine Learning: Science and
Technology} 6 (4): 045067. DOI: 10.1088/2632-2153/ae2382; URL:
\url{https://iopscience.iop.org/article/10.1088/2632-2153/ae2382}.

\end{document}